\documentclass[runningheads]{llncs}
\usepackage[T1]{fontenc}
\usepackage{hyperref}
\usepackage{url}
\usepackage{amsmath}
\usepackage{multirow}
\usepackage{amsfonts}
\usepackage{graphicx,verbatim}
\begin{document}
\title{Anatomy-Aware Low-Dose CT Denoising via Pretrained Vision Models and Semantic-Guided Contrastive Learning}
%

\author{Runze Wang\inst{1,2,3} \and
Zeli Chen\inst{1,2} \and
Zhiyun Song\inst{1,4} \and Wei Fang\inst{1} \and Jiajin Zhang\inst{1,2,4} \and Danyang Tu\inst{1,2} \and Yuxing Tang\inst{1} \and Minfeng Xu\inst{1} \and Xianghua Ye\inst{5} \and Le Lu \inst{1} \and Dakai Jin\inst{1}}
\authorrunning{R. Wang et al.}
%
\institute{DAMO Academy, Alibaba Group \and Hupan Lab, 310023, Hangzhou, China \and Fudan University, Shanghai, China \and Shanghai Jiao Tong University, Shanghai, China \and
The First Affiliated Hospital Zhejiang University, Hangzhou, China 
\\
\email{\{fashe.wrz, dakai.jin\}@alibaba-inc.com}}

\maketitle              
\begin{abstract}

To reduce radiation exposure and improve the diagnostic efficacy of low-dose computed tomography (LDCT), numerous deep learning-based denoising methods have been developed to mitigate noise and artifacts. However, most of these approaches ignore the anatomical semantics of human tissues, which may potentially result in suboptimal denoising outcomes. To address this problem, we propose ALDEN, an anatomy-aware LDCT denoising method that integrates semantic features of pretrained vision models (PVMs) with adversarial and contrastive learning. Specifically, we introduce an anatomy-aware discriminator that dynamically fuses hierarchical semantic features from reference normal-dose CT (NDCT) via cross-attention mechanisms, enabling tissue-specific realism evaluation in the discriminator. In addition, we propose a semantic-guided contrastive learning module that enforces anatomical consistency by contrasting PVM-derived features from LDCT, denoised CT and NDCT, preserving tissue-specific patterns through positive pairs and suppressing artifacts via dual negative pairs. Extensive experiments conducted on two LDCT denoising datasets reveal that ALDEN achieves the state-of-the-art performance, offering superior anatomy preservation and substantially reducing over-smoothing issue of previous work. Further validation on a downstream multi-organ segmentation task (encompassing 117 anatomical structures) affirms the model's ability to maintain anatomical awareness.


\end{abstract}
\section{Introduction}
Low-dose computed tomography (LDCT) has become an important and popular diagnostic tool for reducing radiation exposure risks; however, its clinical utility is hindered by the amplified noise and artifacts that degrade anatomical fidelity. Deep learning advances of convolutional neural networks \cite{chen2023ascon,Zhang2023BoostingSI}, transformer \cite{wang2023ctformer}, and diffusion models \cite{gao2023corediff}, have improved LDCT denoising, these methods share a common limitation: pixel-level constraints (e.g., L1/MSE losses) prioritize global error reduction at the expense of local anatomical plausibility, often resulting in oversmooth textures that obscure small tissues and subtle pathologies \cite{zhang2024review}.

Generative adversarial networks (GANs) offer an alternative by learning data distributions rather than pixel-wise mappings \cite{goodfellow2014generative,wang2022cycmis,wang2024cyclesgan}. However, conventional GAN-based LDCT denoising methods \cite{huang2021gan,wolterink2017generative} often do not capture the important relationship between noise characteristics and anatomical semantics, as noise levels in CT images differ depending on tissue type \cite{mussmann2021organ,chen2023ascon}. Recent work also highlights the need for image understanding in restoration to improve the explainability and clinical application of medical imaging \cite{sun2024coser,li2024sed,chen2023ascon,chen2024low}. This calls for a shift towards fine-grained anatomy-aware denoising, where semantic consistency is crucial for effective texture restoration.

Integrating anatomical semantics into denoising models presents challenges since conventional task-specific segmentation networks \cite{zhang2021task,huang2024lidia} require costly and precise anatomical annotations, limiting generalizability across a large number of diverse anatomies. Progresses in foundation models \cite{oquab2023dinov2,zhang2024disease} demonstrate that pretrained vision models (PVMs) pretrained on large-scale datasets possess exceptional transfer learning capability for semantic understanding. PVMs offer two key advantages: (1) their exposure to millions of natural images allows the development of rich hierarchical feature representations that capture universal texture and structure patterns, which can be adapted to the medical image domain \cite{azizi2023robust,zhao2023clip,ayzenberg2024dinov2,muller2024medical}; and (2) unlike segmentation networks requiring predefined anatomical labels, PVMs generate semantic features without explicit supervision, enabling the discovery of latent anatomical relationships essential for fine-grained denoising.

Inspired by this, we present \textbf{ALDEN} (\textbf{A}natomy-aware \textbf{L}DCT \textbf{DEN}oising framework), which integrates PVMs within a GAN architecture for enhanced anatomy-aware restoration. ALDEN features an anatomy-aware discriminator that utilizes hierarchical semantic features extracted from reference NDCT via PVMs, guiding adversarial learning to concentrate on tissue-specific semantics. This differentiates it from previous GAN-based LDCT denoising methods that evaluate all anatomical structures uniformly. Additionally, we propose a semantically guided contrastive paradigm that uses PVM-extracted features to enforce anatomy-aware consistency. Positive pairs align features from corresponding anatomical regions in denoised CT and NDCT, while negative pairs include features from denoised CT and LDCT at the same location to emphasize noise, and from denoised CT and NDCT at mismatched locations to penalize anatomical misalignment. The InfoNCE loss \cite{chen2020simple} is utilized to minimize the distances between positive pairs and maximize the separation from negative pairs.

Our contributions are summarized as follows. 1) We first propose the integration of PVMs into LDCT denoising, uniquely combining PVMs with adversarial and contrastive learning approaches. 2) We introduce an anatomy-aware discriminator that dynamically incorporates hierarchical semantic features from NDCT to enable a fine-grained semantic-aware LDCT denoising. 3) We present a semantically guided contrastive learning module to maintain anatomical consistency through positive pairs while reducing noise and artifacts with dual negative pairs. 4) Extensive experiments on two LDCT denoising datasets demonstrate that ALDEN achieves the state-of-the-art denoising performance, delivering enhanced texture preservation and avoiding over-smoothing. Further validation on a downstream multi-organ segmentation task with 117 anatomical structures demonstrates the effectiveness of our anatomy-aware denoising approach.

\section{Methodology}

\subsection{Overview of ALDEN} 
\label{Overview of ALDEN}

Conventional GAN-based LDCT denoising frameworks \cite{huang2021gan,wolterink2017generative} employ a generator-discriminator architecture to enhance perceptual quality. Let $X \sim P_X$ denote LDCT inputs and $Y \sim P_Y$ their NDCT counterparts. The generator (i.e., denoising network) $G$ produces denoised outputs $\hat{Y} = G(X)$, optimized through two objectives: 1) \emph{Pixel fidelity} via pixl-wise supervised loss. Here, $L_1$ loss is employed, which is expressed as $\mathcal{L}_1=\| \hat{Y} - Y \|_1$. 2) \emph{Distribution alignment} via adversarial learning. The discriminator $D$ differentiates between real NDCT images $Y$ and generated/denoised outputs $\hat{Y}$. Simultaneously, the generator $G$ is encouraged to produce realistic denoised CT $\hat{Y}$ that can effectively compete with the discriminator. The $G$ and $D$ form a two-player minimax game, optimized through adversarial loss defined as $\mathcal{L}_{adv} = \mathbb{E}_{Y \sim P_Y}[\log D(Y)] + \mathbb{E}_{X \sim P_X}[\log(1 - D(G(X)))]$.

\begin{figure}[t]
\centering
\includegraphics[width=\textwidth]{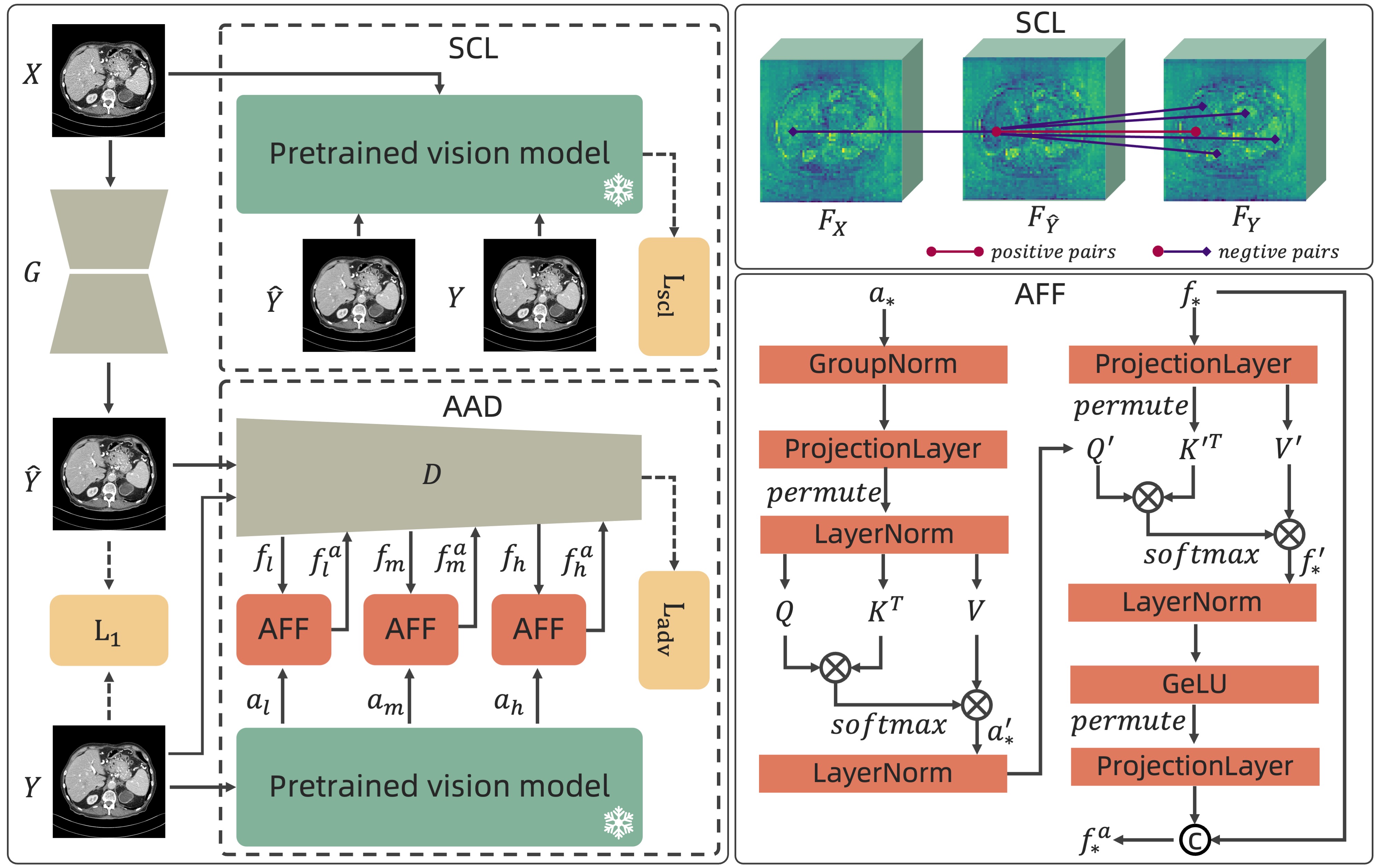}
\caption{Overview of the proposed ALDEN (\textbf{A}natomy-aware \textbf{L}DCT \textbf{DEN}oising) framework, which integrates the Anatomy-Aware Discriminator (AAD) for fine-grained texture restoration and Semantic-guided Contrastive Learning (SCL) for enhanced noise suppression.} \label{method}
\end{figure}

As illustrated in Fig. \ref{method} (left), our ALDEN framework extends the traditional GAN-based LDCT denoising model by introducing two key components: the Anatomy-Aware Discriminator (AAD) and Semantic-guided Contrastive Learning (SCL). The AAD leverages NDCT-derived semantic features extracted from PVMs as conditional input. Let $\Psi $ represent the PVMs, the goal is to align conditional distributions: $P(\hat{Y}|\Psi (Y)) \approx P(Y|\Psi (Y))$, facilitating detailed anatomy-aware texture restoration. In contrast, conventional discriminators operate only on marginal distributions: $P(\hat{Y}) \approx P(Y)$, potentially overlooking essential semantic content. The SCL component utilizes features extracted by PVMs to enforce anatomy-aware consistency through contrastive learning. It achieves this by forming positive pairs to preserve tissue-specific patterns and dual negative pairs to suppress noise and artifacts. More details on AAD can be found in Section \ref{Anatomy-aware discriminator}, while the formalization of SCL is presented in Section \ref{Semantic-guided contrastive learning}.

\subsection{Anatomy-Aware Discriminator}
\label{Anatomy-aware discriminator}

The AAD aims to achieve fine-grained semantic-aware denoising through adversarial learning. Inspired by \cite{li2024sed}, we propose an Attention-based Feature Fusion (AFF) module that integrates hierarchical semantic priors from the reference NDCT $Y$ using PVMs. As shown in Fig. \ref{method} (left), our multi-level AFF operates across the feature hierarchies of both the PVM and the discriminator, providing progressive semantic guidance. To enhance the quality of semantic priors, we also explore advanced PVMs such as DINOv2 \cite{oquab2023dinov2} and MedSAM \cite{ma2024segment}, which are the state-of-the-art pretrained vision models in the fields of natural and medical images, offering robust semantic feature extraction capabilities.

Given the NDCT $Y$, we extract hierarchical semantic features from a fixed PVM at three levels: low ($a_l$), middle ($a_m$), and high ($a_h$). Both MedSAM and DINOv2 use the ViT-base architecture with 12 transformer blocks. Although ViT doesn’t explicitly change scales, features from different layers are analogous to CNNs: early layers focus on low-level features, while later ones concentrate on high-level semantics \cite{dosovitskiy2020image}. We use outputs from the 4th, 8th, and 12th transformer blocks as low-, mid-, and high-level features, respectively. At the same time, when the denoised CT $\hat{Y}$ or the NDCT $Y$ is the input to the discriminator $D$, it generates the corresponding hierarchical discriminative features, i.e., $f_{l}$, $f_{m}$ and $f_{h}$ at these levels. Then multiple AFF modules are used to align the features $f_{*}$ (where $* \in {l, m, h}$) with the semantic features $a_{*}$ before passing them to the discriminator. This process guides the discriminator to focus on semantically relevant texture distributions.

As shown in Fig. \ref{method} (bottom right), the AFF module begins by standardizing the semantic features $a_{*}$ through group normalization, followed by a projection layer with a $1*1$ convolution. After permutation and layer normalization, we derive the query $Q$, key $K$, and value $V$. The self-attention outputs $a'_{*}$ are computed as:
\begin{equation}
\label{equation:attention}
    a'_{*} = Softmax(Q\cdot K^{T}/\sqrt{d_{k}} )\cdot V,
\end{equation}
where $d_{k}$ is the scale factor. This output $a'_{*}$ is normalized to produce the query $Q'$ for the cross-attention mechanism. The feature $f_{*}$ undergoes a similar projection and permutation process to become the key $K'$ and the value $V'$. Then we can obtain the cross-attention results $f'_{*}$ using Equation \ref{equation:attention}. Finally, the fused anatomy-aware feature representation $f_{*}^{a}$ is computed as follows:
\begin{equation}
    f_{*}^{a} = Concat(PL(Permute(GELU(LN(f_{*}')))), f_{*}).
\end{equation}
Here, $GELU$, $PL$, and $LN$ are the Gaussian Error Linear Units, projection layer and layer normalization, respectively. 

\subsection{Semantic-guided Contrastive Learning}
\label{Semantic-guided contrastive learning}

The SCL component of the ALDEN framework utilizes a pretrained vision model to improve LDCT denoising by ensuring anatomical consistency between denoised CT and reference NDCT. As depicted in Fig. \ref{method} (top right), the feature representation $F_{X}$, $F_{\hat{Y}}$ and $F_{Y}$ are derived from the fixed PVM using input $X$, $\hat{Y}$ and $Y$, respectively. Here, $F_{X}$, $F_{\hat{Y}}$ and $F_{Y} \in \mathbb{R}^{B\times C\times H\times W}$, where $B$, $C$, $H$ and $W$ represent batch size, channel depth, height, and width of the features, respectively. Then, SCL operates contrastive learning on these features, explicitly aligning denoised CT features with NDCT references while contrasting against two types of negative samples: residual noise patterns from LDCT and anatomically discordant NDCT features.

\noindent\textbf{Positive Pair Alignment.} For each denoised CT image, we establish anatomical correspondence with its NDCT counterpart through spatially aligned feature pairs. Let $F_{\hat{Y}}(x,y)\in \mathbb{R}^C$ and $F_{Y}(x,y)\in \mathbb{R}^C$ denote the PVM-derived feature vectors at the spatial coordinate $(x,y)$ in the $i$-th batch sample. We randomly sample $K$ coordinates per image, constructing positive pairs from identical anatomical locations, i.e., $\mathcal{P}=\{(F_{\hat{Y}}^{(i)}(x_{ik},y_{ik})), F_{Y}^{(i)}(x_{ik},y_{ik})\}_{i=1,k=1}^{B,K}$, where $(x_{ik},y_{ik})$ represents randomly selected positions. This alignment ensures that the denoised output preserves anatomical structures observed in the NDCT ground truth.

\noindent\textbf{Dual Negative Sampling Strategy.} We design two complementary negative sampling mechanisms: 1) same-location LDCT negatives capture residual noise by contrasting denoised features against their LDCT counterparts at identical coordinates, i.e., $\mathcal{N}_1=\{(F_{\hat{Y}}^{(i)}(x_{ik},y_{ik})), F_{X}^{(i)}(x_{ik},y_{ik})\}_{i=1,k=1}^{B,K}$. 2) Cross-location NDCT negatives penalize anatomical misalignment by pairing denoised features with NDCT features from spatially discordant regions $M$, i.e., $\mathcal{N}_2=\{(F_{\hat{Y}}^{(i)}(x_{ik},y_{ik})), F_{Y}^{(i)}(\tilde{x}_{ik}^{(m)},\tilde{y}_{ik}^{(m)})\}_{i=1,k=1,m=1}^{B,K,M}$, where $\{(\tilde{x}_{ik}^{(m)},\tilde{y}_{ik}^{(m)})\}$ are random-sampled coordinates excluding $(x_{ik},y_{ik})$. 

\noindent\textbf{Loss Formulation.} The SCL loss adapts the InfoNCE \cite{chen2020simple} to optimize feature similarities:
\begin{equation}
    \mathcal{L}_{scl} = -\frac{1}{B \cdot K} \sum_{i, k} \log \frac{exp(s_{ik}^{pos}/ \tau)}{exp(s_{ik}^{pos}/ \tau) + exp(s_{ik}^{neg_1}/ \tau)+  {\textstyle \sum_{m=1}^{M}} exp(s_{ikm}^{neg_2}/ \tau)},
\end{equation}
where the temperature $\tau=0.1$ and the similarity scores are computed as $s_{ik}^{pos}=\left \langle F_{\hat{Y}}^{(i)}(x_{ik},y_{ik}), F_{Y}^{(i)}(x_{ik},y_{ik})\right \rangle$, $s_{ik}^{neg_1}=\left \langle F_{\hat{Y}}^{(i)}(x_{ik},y_{ik}), F_{X}^{(i)}(x_{ik},y_{ik})\right \rangle$ and $s_{ikm}^{neg_2}=\left \langle F_{\hat{Y}}^{(i)}(x_{ik},y_{ik}), F_{Y}^{(i)}(\tilde{x}_{ik}^{(m)},\tilde{y}_{ik}^{(m)})\right \rangle$, respectively. Here, $\left \langle\cdot,\cdot\right \rangle $ denotes cosine similarity, and the loss simultaneously maximizes positive pair alignment while repelling both negative types. The overall objective function of the proposed ALDEN is as follows:
\begin{equation}
\mathcal{L}=\mathcal{L}_1+\lambda_1\mathcal{L}_{adv}+\lambda_2\mathcal{L}_{scl},
\end{equation}
where $\{\lambda_1,\lambda_2\}$ are parameters controlling the relative weights of different losses, which are empirically set as $\{0.01, 0.5\}$.

\section{Experiments and Results}

\subsection{Experimental Setup}
\noindent\textbf{Denoising Datasets.} We evaluate our method using two datasets. Mayo2016 dataset \cite{mccollough2017low} consists of 2,378 CT slices from ten anonymized patients, each of which has paired low-dose (quarter-dose) and normal-dose scan. Consistent with the data split protocol of previous studies \cite{gao2023corediff,wang2023ctformer}, we select slices from nine patients for training and slices from one patient for testing. The second dataset is a collection of CT scans from multiple centers, referred to as the Multi-center CT Denoising (MCTD) dataset. Normal-dose CT scans were acquired from diverse clinical sites, while the corresponding low-dose CT scans were generated using a simulation algorithm \cite{yu2012development}. The dataset consists of 1,276 paired scans for training and 88 paired scans for validation. For both datasets, 2D slices were extracted from the axial plane with a resolution of 512×512 pixels for model training and validation.

We employ four metrics for denoising evaluation: peak signal-to-noise ratio (PSNR), structural similarity index measure (SSIM), root mean square error (RMSE) and a perceptual metric, i.e., learned perceptual image patch similarity (LPIPS) \cite{zhang2018unreasonable}, to assess image quality.

\noindent\textbf{Segmentation Dataset.} We further evaluate the denoising performance on a downstream multi-organ segmentation task. Specifically, the TotalSegmentator \cite{wasserthal2023totalsegmentator} test set is used, which includes 89 CT scans with 117 organ types. To simulate LDCT data with varying noise levels, we follow the methodology in \cite{yu2012development} to generate LDCT with low and high noise conditions. Then different denoising algorithms are applied to restore the simulated LDCT, after which we use the nnUNet~\cite{isensee2021nnu} trained on the original CT in the TotalSegmentator training set to predict organ masks and calculate the mean Dice similarity coefficient (DSC) of all organs.

\noindent\textbf{Implementation Details.} The generator utilized in our study is the ESAU-Net as introduced in \cite{chen2023ascon}. The basic discriminator we used is a popular patch-wise discriminator \cite{isola2017image} that consists of five convolutional layers. The proposed AAD incorporates the AFF within the middle three convolutional layers. The sampling hyperparameters $K$ and $M$ in the SCL module are empirically set to 256 and 32, respectively. Our experiments are conducted on an NVIDIA H20 GPU with 96 GB of memory using the PyTorch framework. We optimize our network with the Adam optimizer, utilizing a batch size of 8 and a learning rate of 1e-4. The optimization process is carried out for a total of 300,000 iterations.

\subsection{Experimental Results}

\noindent\textbf{Comparison with Previous State-of-the-Art Methods.}  We implement two ALDEN variants based on different PVMs: ALDEN-MedSAM and ALDEN-DINOv2. These variants are compared against six state-of-the-art denoising methods with diverse network architectures, including GAN-based (DU-GAN \cite{huang2021gan} and SeD \cite{li2024sed}), CNN-based (RED-CNN \cite{chen2017low} and ASCON \cite{chen2023ascon}), Transformer-based (CTformer \cite{wang2023ctformer}) and Diffusion-based (CoreDiff \cite{gao2023corediff}).

\begin{table}[tbp]
    \centering
    \caption{Quantitative comparison of different methods on the Mayo2016 and MCTD datasets.}
    \begin{tabular}{l|cccc|cccc}
    \hline
    \multirow{2}{*}{Methods} & \multicolumn{4}{c|}{Mayo2016 dataset} & \multicolumn{4}{c}{MCTD dataset} \\ \cline{2-9}
                             & PSNR$\uparrow$ & SSIM$\uparrow$ & RMSE$\downarrow$ & LPIPS$\downarrow$ & PSNR$\uparrow$ & SSIM$\uparrow$ & RMSE$\downarrow$ & LPIPS$\downarrow$ \\ \hline
    RED-CNN \cite{chen2017low} & 32.32 & 0.9227 & 9.87 & 0.0235 & 37.81 & 0.9069 & 50.11 & 0.0535 \\
    DU-GAN \cite{huang2021gan} & 32.64 & 0.9278 & 9.56 & 0.0198 & 38.52 & 0.9075 & 43.33 & 0.0291 \\
    SeD \cite{li2024sed} & 32.87 & 0.9283 & 9.32 & 0.0201 & 39.43 & 0.9157 & 40.78 & 0.0304 \\ 
    CTformer \cite{wang2023ctformer} & 33.16 & 0.9287 & 8.98 & 0.0235 & 38.97 & 0.9134 & 40.54 & 0.0458 \\ 
    CoreDiff \cite{gao2023corediff} & 33.61 & 0.9342 & 8.58 & 0.0227 & 40.46 & 0.9290 & 34.40 & 0.0474 \\ 
    ASCON \cite{chen2023ascon} & 33.60 & 0.9318 & 8.57 & 0.0242 & \textbf{40.58} & 0.9278 & 34.43 & 0.0657 \\ \hline
    ALDEN-MedSAM & 33.59 & \textbf{0.9343} & 8.56 & 0.0192 & 40.51 & 0.9281 & 34.86 & 0.0273 \\
    ALDEN-DINOv2 & \textbf{33.71} & 0.9341 & \textbf{8.44} & \textbf{0.0176} & 40.57 & \textbf{0.9296} & \textbf{34.37} & \textbf{0.0265} \\  \hline
    \end{tabular}
    \label{tab:comparison}
\end{table}

Table \ref{tab:comparison} reveals an inherent trade-off between fidelity metrics (PSNR, SSIM, RMSE) and perceptual quality (LPIPS) in previous work. For example, on the Mayo2016 dataset, DU-GAN achieves a competitive LPIPS score of 0.0198 but underperforms in fidelity with an RMSE of 9.56, compared to 8.57-8.58 for ASCON and CoreDiff. Conversely, CoreDiff and ASCON exhibit superior fidelity (PSNR: 33.60-33.61 dB) but have higher LPIPS values (0.0227-0.0242), indicating a decline in perceptual quality. In contrast, the ALDEN variants, especially ALDEN-DINOv2, effectively balance these objectives. From the Mayo2016 dataset, ALDEN-DINOv2 achieves a new state-of-the-art result, reporting a PSNR of 33.71 dB, an RMSE of 8.44, and an LPIPS of 0.0176, along with a competitive SSIM of 0.9341. In MCTD dataset, it maintains its superiority, leading in SSIM (0.9296), RMSE (34.37) and LPIPS of 0.0265, indicating a 44.1-59.7\% reduction compared to CoreDiff and ASCON. These results confirm ALDEN's ability to harmonize fidelity and perceptual quality through PVM guidance.
\begin{figure}[tbp]
    \centering
    \includegraphics[width=\linewidth]{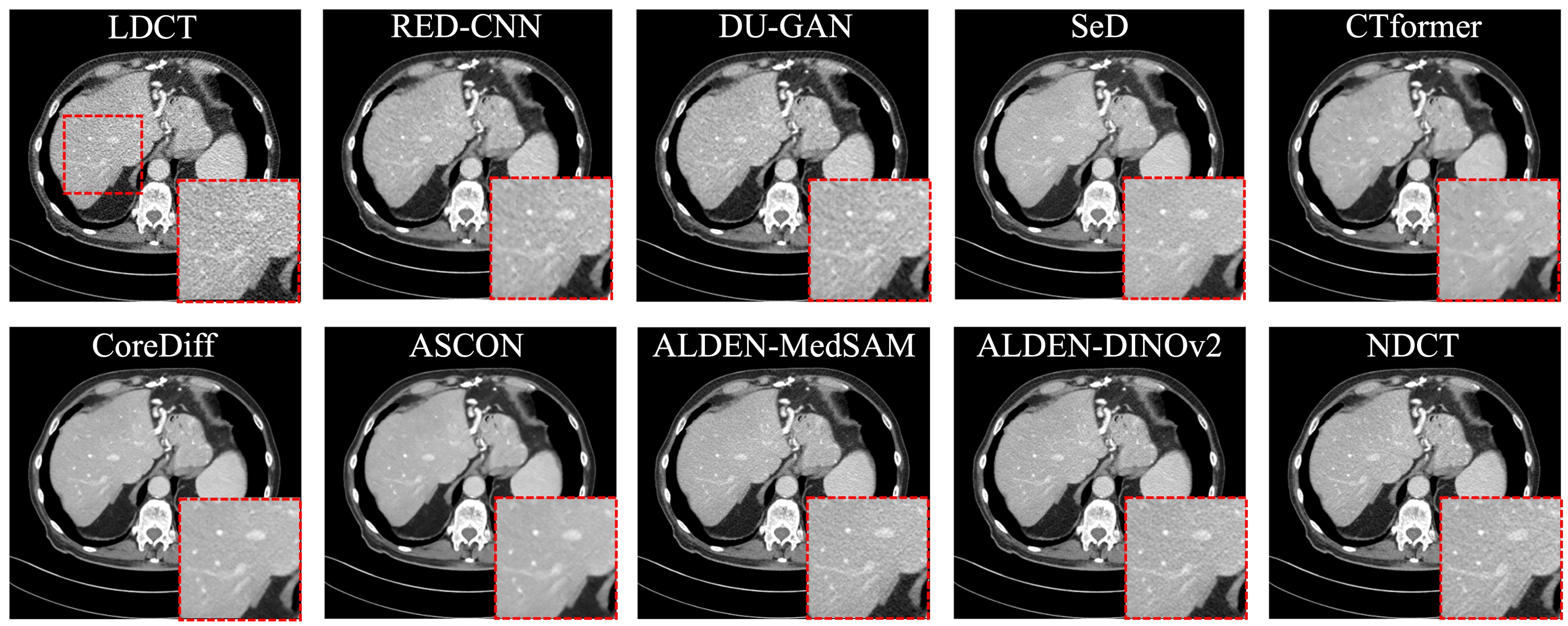}
    \caption{Qualitative assessment of different methods on the Mayo2016 dataset. The display window is $[-160,240]$ HU.}
    \label{fig:qualitative_results} 
    \vspace{-1mm}
\end{figure}
Fig. \ref{fig:qualitative_results} presents the qualitative results of various methods applied to the Mayo2016 dataset. As shown, the proposed ALDEN-MedSAM and ALDEN-DINOv2 stand out by effectively preserving intricate textural details while maintaining exceptional noise reduction, leading to results that closely resemble the NDCT image.


\noindent\textbf{Ablation Studies.} Table \ref{tab:ablation} showcases the ablation results of ALDEN-DINOv2 in the Mayo2016 dataset, highlighting the roles of AAD and SCL in improving the performance of LDCT denoising. It is observed that incorporating AAD notably enhances PSNR, SSIM, RMSE and LPIPS as compared to the baseline model. Applying SCL independently produces improvements in these metrics, particularly in LPIPS, likely due to PVMs’ efficacy as perceptual feature extractors. The combination of AAD and SCL delivers the best performance, achieving a PSNR of 33.71, RMSE of 8.44, and maintaining balanced results in SSIM (0.9341) and LPIPS (0.0176), highlighting the effectiveness of integrating these components to enhance denoising fidelity and maintain perceptual quality.
\begin{table}[tbp]
    \centering
    \caption{Ablation study results on the Mayo2016 dataset.}
    \resizebox{0.65\textwidth}{!}{
    \begin{tabular}{l|cc|cccc}
        \hline
        Methods & AAD & SCL & PSNR$\uparrow$ & SSIM$\uparrow$ & RMSE$\downarrow$ & LPIPS$\downarrow$ \\
        \hline
        Baseline & - & - & 32.70 & 0.9291 & 9.51 & 0.0206 \\
        \hline
        AAD-DINOv2 & $\checkmark$ & - & 33.60 & \textbf{0.9348} & 8.57 & 0.0186 \\
        SCL-DINOv2 & - & $\checkmark$ & 33.41 & 0.9330 & 8.74 & \textbf{0.0123} \\
        ALDEN-DINOv2 & $\checkmark$ & $\checkmark$ & \textbf{33.71} & 0.9341 & \textbf{8.44} & 0.0176 \\
        \hline
    \end{tabular}
    }
    \label{tab:ablation}
\end{table}
\begin{table}[tbp]
\centering
\caption{Quantitative comparison of different denoising methods on DSC (\%) for downstream multi-organ segmentation tasks.}
\label{tab: segmentation}
\resizebox{\textwidth}{!}{
\begin{tabular}{l|cccccccc}
\hline
Noise level & LDCT & RED-CNN  & DU-GAN & SeD & CTformer & CoreDiff & ASCON  & ALDEN-DINOv2   \\ \hline
Low & 87.54 & 88.22 & 88.74 & 88.68 & 88.53 & 89.06  & 88.94  & 89.20 \\
High & 75.74 & 78.14 & 78.01 & 79.37 & 78.96 & 79.99 & 79.60 & 81.06  \\ \hline
\end{tabular}
}
\end{table}

\noindent\textbf{Downstream Task Evaluation.}
We evaluate the performance of the proposed ALDEN-DINOv2 for multi-organ segmentation as downstream task.  As shown in Table \ref{tab: segmentation}, our method consistently achieves the best segmentation performance in both low- and high-noise scenarios, with DSC values of 89.20\% and 81.06\%, respectively. Especially in the high-noise scenario, our method substantially outperforms the second-best CoreDiff by a mean DSC of 1.07\% across 117 anatomical structures. These results demonstrate that our approach effectively improves anatomical perception and enhances segmentation performance.


\section{Conclusion}
In conclusion, we present ALDEN that combines pretrained vision models with adversarial and contrastive learning techniques for anatomy-aware low-dose CT denoising. The framework features an anatomy-aware discriminator for fine-grained denoising and a semantic-guided contrastive learning module to enhance anatomical consistency. Extensive experiments demonstrate that ALDEN achieves state-of-the-art performance, improving texture preservation while reducing over-smoothing. Validation on a multi-organ segmentation task with 117 anatomical structures underscores the model's robust anatomical awareness.

\begin{credits}

\subsubsection{\discintname}
The authors have no competing interests to declare that are relevant to the content of this article.
\end{credits}
%
%
%
\bibliographystyle{splncs04}
\bibliography{paper-2266}
%




\end{document}